\documentstyle{mn}
\input{psfig.sty}
\newcommand{\beq}{\begin{equation}}
\newcommand{\eeq}{\end{equation}}
\newcommand{\barray}{\begin{eqnarray}}
\newcommand{\earray}{\end{eqnarray}}

\newcommand{\bitem}{\begin{itemize}}
\newcommand{\eitem}{\end{itemize}}

\newcommand{\mpc}{\,{\rm {Mpc}}}
\newcommand{\mpch}{\,h^{-1}{\rm {Mpc}}}
\newcommand{\kpc}{\,{\rm kpc}}

\newcommand{\kms}{\,{\rm {km\, s^{-1}}}}
\newcommand{\msun}{{\,\rm M_\odot}}
\newcommand{\Lsun}{{\,\rm L_\odot}}

\newcommand{\Omnow}{\Omega_{\rm m,0}}

\def\reference{\bibitem[]{}}
\def\md{m_{\rm d}}
\def\Mh{M_{\rm h}}
\def\Md{M_{\rm d}}

\def\cc{c^\prime}
\def\cs{c}
\def\Eorb{E_{\rm orb}}
\def\Ms{M_{\rm s}}
\def\rh{r_{\rm h}}
\def\Mh{M_{\rm h}}
\def\rc{r_{\rm c}}
\def\zf{z_{\rm f}}
\def\rs{r_{\rm s}}

\newdimen\hssize
\newdimen\hdsize 
\begin{document}

\title[Galaxy formation in pre-processed dark halos]
{Galaxy formation in pre-processed dark halos}

\author[Mo and Mao]
{H.J. Mo$ ^{1}$ and Shude Mao$ ^{2}$
\thanks{E-mail: hjmo@nova.astro.umass.edu; smao@jb.man.ac.uk}\\
$^{1}$ Department of Astronomy, University of Massachusetts, 
       Amherst, MA 01002, USA\\
$^{2}$ Jodrell Bank Observatory, Macclesfield, Cheshire SK11 9DL, UK}

\date{}

\maketitle

\label{firstpage}

\begin{abstract}
 Recent $N$-body simulations show that the formation of a present-day, 
galaxy sized dark matter halo in the cold dark matter cosmogony
in general consists of an early fast collapse phase,
during which the potential associated with a halo is established,
followed by a slow accretion phase, during which mass 
is added rather gently in the outer region.  
In this paper, we consider the implication of such 
a halo assemble history for galaxy formation. We outline a scenario 
in which the fast collapse phase is accompanied with 
rapid formation of cold clouds and with starbursts 
that can eject a large amount of gas from the halo centre.
The loss of orbital energy of the cold clouds to the dark matter 
and the ejection of gas from the halo centre by 
starbursts can significantly reduce the halo concentration. 
The outflow from the starburst can also heat the gas in the 
protogalaxy region. Subsequent formation of galaxies in the 
slow accretion regime is therefore in halos that have been 
pre-processed by these processes and may have properties 
different from that given by $N$-body simulations.
This scenario can help to solve several outstanding problems 
in the standard $\Lambda$CDM model of galaxy formation
without compromising its success in allowing structure
formation at high redshift. The predicted rotation curves 
can be significantly flatter than those based on 
the halo profiles obtained from $N$-body simulations, 
alleviating the discrepancy of the Tully-Fisher 
relation predicted in the standard $\Lambda$CDM model
with observations. The flattened galaxy halos allow 
accreted minihalos to survive in their central regions longer,
which may be helpful in producing the flux anomalies 
observed in some gravitational lensing systems. The preheating 
by the early starbursts effectively reduces the amount of gas that can be 
accreted into galaxy halos, which may explain why the baryon fraction
in a spiral galaxy is in general much
lower than the universal baryon fraction, 
$f_B\sim 0.16$, in the standard $\Lambda$CDM model.
\end{abstract}


\section{Introduction}

For many years, the study of galaxy formation has to 
contend with uncertainties both in cosmology 
(which determines the spacetime background and initial 
conditions for structure formation) and in 
the physical processes that drive evolution.
Thanks to recent observations on the Cosmic Microwave 
Background (CMB) anisotropy and large-scale structure,
the cosmological uncertainties are largely resolved, and the 
study of galaxy formation can now focus on the physical processes.    
There is now much evidence that we live in a 
flat universe which is dominated by Cold Dark Matter 
(CDM), with a total energy density $\Omega_{0}=1.02\pm 0.02$, 
total matter density $\Omnow = (0.135\pm 0.009)h^{-2}$, 
baryon density $\Omega_{\rm B,0}= (0.0224\pm 0.0009) h^{-2}$, 
Hubble constant (in units of $100\kms\mpc^{-1}$) 
$h=0.71\pm 0.04$, the power-law index of initial perturbation
$n\sim 1$, and the amplitude of the perturbation power 
spectrum, as specified by the {\it rms} of the 
perturbation field smoothed in spheres of a radius 
$8\mpch$, $\sigma_8=0.85\pm 0.1$. The values quoted here are based 
on the recent results obtained from WMAP observations of the CMB, 
but are consistent with a large number of other 
observations (e.g. Spergel et al. 2003). Assuming the density 
perturbation to be Gaussian,  this (small) set of parameters 
completely determines the CDM model for structure formation. 
Thus, although the nature of the dark matter and the 
cosmological constant remains to be explained, there is not 
much freedom in choosing the spacetime background and initial conditions
for galaxy formation.

  When combined with a set of 
standard (simple) assumptions about galaxy 
formation, this cosmogony has several outstanding problems.
First, $N$-body simulations have constantly shown that  
dark matter halos in this cosmogony are quite concentrated,
which seems to be at odds with the observed rotation curves of dwarf 
galaxies (e.g. Moore 1994; 
Burkert 1995; van den Bosch et al. 2000; 
de Blok, McGaugh \& Rubin 2001; Borriello \& Salucci 2001; 
Blais-Ouelette, Amram \& Carignan 2001; Dutton et al. 2003)
and the observed 
Tully-Fisher relation for normal disk galaxies  
(Mo \& Mao 2000; Navarro \& Steinmetz 2000; van den Bosch 
et al. 2003). Because of this, alternative models are proposed,
where either the dark matter is assumed to be
warm or collisional, or the initial power spectrum is assumed to be
strongly tilted (e.g. Spergel \& Steinhardt 2000; Hogan \& Dalcanton
2000; Kamionkowski \& Liddle 2000).  
Although these models may help to 
reconcile some of the problems outlined above, they are not 
favored by the recent WMAP result that the formation of 
the first generation of ionizing sources may have occured 
at quite high redshift. The other problem is an old one. 
When modern CDM dominated theories of 
galaxy formation were first considered
in the late seventies (e.g. White \& Rees 1978), 
it was immediately realized that
some processes must be invoked to prevent gas from cooling too
fast, otherwise all the gas would cool and 
turn into stars by the present day.
Although various feedback schemes
have been proposed (Dekel \& Silk 1986) 
to solve this `overcooling' problem, a satisfactory solution  
remains lacking. In the standard $\Lambda$CDM model
favored by current observations, the problem is much more severe 
since the baryon fraction is approximately $16\%$, 
which is 3 times higher than that in the standard CDM model
with $\Omega_{\rm m}=1$, making radiative  cooling even more effective. 
The observed baryonic mass fraction in present-day galaxy
halos is only a few percent (see \S\ref{ssec_bf}); this poses another
serious problem for the present cosmogony , because one has to explain why only such a small fraction of
the gas in a protogalactic region eventually assembles into a galaxy.
The conventional feedback scheme, where energy feedback from supernova
explosions is assumed to drive the cooled gas in a formed galaxy disk
out of a galactic halo, may not be a viable solution, because such
feedback is only effective in very low mass galaxies (e.g. Mac Low \& Ferrara
1999).

The existence of these problems is not necessarily a failure of the
standard $\Lambda$CDM model because galaxy formation involves many
complex physical processes in addition to the cosmological initial and
boundary conditions. It is, therefore, possible that we are indeed
living in a $\Lambda$CDM universe, but that some important pieces of
the puzzle are still missing in our current understanding of galaxy
formation.  In particular, the dynamical role of the baryon component
and star formation in structure formation are two important aspects of
the problem that have not been fully understood.  Since the universal
baryon fraction in current cosmogony is quite high and since baryonic
gas is dissipational, the baryon component may have played an
important role in structure formation, especially on galactic scales.
In this paper, we examine the potential of such dynamical
effects in solving the problems discussed above.

The scenario we are proposing is motivated by the following
considerations. According to recent $N$-body simulations
(e.g. Wechsler et al. 2002; Zhao et al. 2003a,b), the formation of
a galaxy halo in the $\Lambda$CDM cosmogony typically 
consists of two distinct phases:
a fast collapse phase during which the gravitational potential
associated with a halo is established, and a slow-accretion phase
during which mass is added gently to the halo without significantly
affecting the potential well. 
The early fast-collapse phase is dominated by
violent mergers of dark matter halos. Since radiative cooling 
of gas is very effective in galaxy-sized halos at high redshift
(with cooling time much shorter than halo collapse time),
this phase is expected to be accompanied by rapid formation of 
dense gas clouds, which then rapidly collapse into
the centre of the galaxy halo. 
During this process, the structure of dark matter halos can be
affected by the baryonic component.  The violent collapse of the gas
clouds in the fast-collapse phase may also trigger an episode of
rapid star formation that may drive a large amount of gas out from the
halo centre. Such mass loss will cause the halo to expand in the inner
region, thereby reducing the halo concentration. 
As we will show below,  these two processes combined
may be sufficient to overcome the concentration problem in 
the standard $\Lambda$CDM model, given the high baryonic mass 
fraction in this cosmogony. This early episode of starbursts 
associated with the fast-collapse phase may also heat the gas 
in a protogalaxy before it collapses, thereby reducing the amount 
of gas that can be accreted during the slow-accretion phase. 
A proper treatment of these processes
is therefore also a key step in understanding the baryon fraction 
in normal galaxies
(e.g. Mo \& Mao 2002; Oh \& Benson 2003; Granato et al 2003).  

 Some of these issues we are considering here have already been 
discussed in earlier investigations. 
The importance of the baryon component in affecting
halo profile was pointed out in Binney et al. (2001); El-Zant et al. (2001) 
considered the possibility that the interaction between gas 
clouds and dark halo can erase the cusps in the halo of dwarf 
galaxies; Weinberg \& Katz (2002) found that the secular   
evolution of a bar-like structure in a dark matter 
halo can effectively get rid of the cusps of CDM halos;
Navarro et al. (1996) considered the possibility that a sudden blowout
of baryons from the centre of a dwarf galaxy can cause 
significant flattening of the inner profile of its halo.
The effect of preheating on subsequent gas cooling and galaxy 
formation has been considered in Mo \& Mao (2002) and in 
Oh \& Benson (2003). Our study here intends to present a 
related discussion in the context of dark halo formation 
in the current CDM cosmogony, and in particular, we want to 
demonstrate that a proper treatment of these processes 
may be pivotal in solving many of the vexing problems in 
current models of galaxy formation.  

  This paper is organized as follows. In Section 2 we revisit
some of the problems about galaxy formation in the standard 
$\Lambda$CDM model. In Section 3 we examine 
two possible ways in which the baryon component can affect halo 
concentration, and we discuss the observational consequences 
of the change of halo concentrations in Section 4. 
In Section 5, we examine how an early phase of starburst, which may 
be responsible for the formation of bulge and halo stars, 
can affect the subsequent formation of galaxies in dark halos. 
Finally, in Section 6, we make further discussion about our results.

\section {Some outstanding problems of galaxy formation
in the standard $\Lambda$CDM model}

\subsection{The properties of CDM halos} 
\label{DH_properties}

 One important aspect of galaxy formation in the CDM cosmogony
is the formation of dark matter halos.
Since dark matter particles take part only in gravitational
interaction, $N$-body simulations can be used to study in detail
the formation of dark halos in the cosmic density field,
and a great deal has been learned about the properties of the
halo population in the past few years.
One important finding is that CDM halos can be 
approximated by a universal profile of the form
\beq
\rho(r) = {4\rho_s\over (r/\rs)(1+r/\rs)^2}\,,
\eeq  
where $\rs$ is a scale radius and $\rho_s$ is the density
at this radius (Navarro et al. 1997, hereafter NFW;
see Moore et al. 1998, Jing 2000, Klypin et al. 2001
for discussions about the uncertainty of the inner 
profile).  
The proper size of a halo is often defined so that the mean density 
within the halo radius $\rh$ is a factor $\Delta_{\rm h}$ times
the mean density of the universe, $\overline {\rho}$, at the
redshift $z$ in consideration. The halo mass $\Mh$
is then related to the halo radius by 
\beq
\Mh={4\pi\over 3} {\overline\rho} \Delta_{\rm h} \rh^3\,.
\eeq
In our discussion we adopt 
$\Delta_{\rm h}=\Delta_{\rm vir}=(18\pi^2+82x-39x^2)/\Omega_{\rm m}(z)$,
where $x=\Omega_{\rm m}(z)-1$, and $\Omega_{\rm m}(z)$ is the   
mass density parameter at redshift $z$
(Bryan \& Norman 1998). For the standard $\Lambda$CDM model, 
$\Delta_{\rm h}\approx 340$
at $z=0$. The parameter that characterizes the concentration of mass
distribution in a halo is the concentration parameter,
defined to be $\cs\equiv \rh/\rs$. It is then easy to show that
the total mass within a radius $r$ is
\beq
\Mh(<r)=\Mh f(\cs x)/f(\cs), 
\eeq
where $x\equiv r/\rh$, and  
\beq
f(x)\equiv \ln (1+x)-x/(1+x)\,.
\eeq
The mass within $\rs$ is
\beq \label{eq:Ms}
\Ms={\ln 2-1/2\over f(\cs)}\Mh\,.
\eeq
Defining the circular velocity of a halo as 
$V_{\rm h}=(G\Mh/\rh)^{1/2}$, we have 
\beq\label{MhasVh}
\Mh={V_{\rm h}^3\over [\Delta_{\rm h}\Omega_{\rm m}(z)/2]^{1/2}G H(z)}\,,
\eeq
where $H(z)$ is the Hubble constant at $z$. 
The gravitational potential of the halo at a radius
$r$ from the centre is  
\beq\label{Phiasr}
\Phi (r)=-V_{\rm h}^2
{1\over f(\cs)} {\ln (1+\cs x)\over x}\,,
\eeq
and the escaping velocity is 
$V_{\rm esc}(r)=\sqrt{-2\Phi(r)}$.

With large $N$-body simulations, it is found that 
the typical value of the concentration depends on halo mass. 
For the standard $\Lambda$CDM model, this mass dependence
can be parameterized as  
\beq\label{casM}
\cs(M)=11 \left({\Mh\over 10^{12}h^{-1}\msun}\right)^{0.15}
\eeq
for halos identified at $z=0$
(e.g. Bullock et al. 2001; Zhao et al. 2003b). 
The value of $\cs$ also depends on 
redshift. Numerical simulations show that the concentration 
of a halo is tightly correlated with its mass accretion history
(Wechsler et al. 2002; Zhao et al. 2003a,b). 
As shown in Zhao et al. (2003a), the formation of each 
CDM halo in general consists of two phases: 
a fast collapse phase during which the gravitational potential 
is established (the formation of the core of potential)
and a slow-accretion phase during which 
mass is gently added  to the core of potential 
without significantly affecting the potential well. 
The formation of the core of potential is dominated by rapid 
mergers and halos in this phase 
have relatively low concentration, $\cs\sim 4$. In the 
slow-accretion phase, the values of inner quantities,
such as $\rs$, $\Ms$ and $V_s$, change little, while 
$\cs$ increases as more mass is added in the outer part of
the halo. Based on this, one may define a formation time 
$t_f$ at which the circular velocity as a function of time
remains roughly constant. Zhao et al. found that the 
formation time defined in this way is closely related to the 
concentration parameter, and for present-day halos
the correlation is well described by   
\beq
\cs\approx 4\times \left[{H(\zf)\over H_0}\right]^{1/\eta}\,,
\eeq
where $\eta\sim 1$. 
We can invert this relation to write
the formation time as a function of $\cs$. 
For a flat universe, we have
\beq
1+\zf\approx \left[\left({\cs\over 4}\right)^{2\eta}
-\Omega_\Lambda\right]^{1/3}
\Omega_{\rm m,0}^{-1/3}\,.
\eeq
For halos of a given mass (or a given circular velocity),
the median concentration defines a typical formation 
time $t_f$, while 
the scatter in $\cs$ implies a scatter in $t_f$.
The median of $\cs$ for a $10^{12}h^{-1}\msun$ halo 
at the present time is about 11, corresponding to a 
formation redshift $\zf=2$. 
For a given mass, the dispersion in $\cs$ is about 
40\%, which corresponds to a formation redshift 
between 1 and 3 for present-day halos with 
$\Mh\sim 10^{12}h^{-1}\msun$.
  
\subsection{The concentration problem}

 It has been pointed out by several authors that the cuspy
NFW halos are at odds with the observed rotation curves of low-surface 
brightness galaxies, which are fit better by profiles 
with a constant density core (e.g. de Blok et al. 2001). 
Even if one enforces the 
NFW profile in the fit, the obtained concentration 
parameters are usually much lower than that predicted 
by the standard $\Lambda$CDM model (e.g. Moore 1994; 
Burkert 1995; Borriello \& Salucci 2001; Blais Ouelette, 
Amram \& Carignan 2001). Although there have been concerns
over the effects of beam smearing (van den Bosch
et al. 2000; Swaters, Madore \& Trewhella 2000), 
recent analyses of high-resolution ${\rm H\alpha}$-HI rotation curves 
(de Blok, McGaugh \& Rubin 2001; and McGaugh, 
Barker \& de Blok 2003) suggest that the discrepancy
appears to be real. Similar problems may exist also for  
normal spiral galaxies. Matching model predictions 
with the observed Tully-Fisher relation, Mo \& Mao (2000)
find that $\cs\sim 4$ is required to obtain an agreement.
The constraint here comes from the fact that 
a higher halo concentration leads to a larger boost
of the disk maximum rotation velocity relative to the halo 
circular velocity, and too high a concentration
leads to too low a luminosity for a given maximum rotation 
velocity. In what follows, we re-visit this problem with the 
use of more recent data, putting it in the context 
of our discussions in the present paper.   

\subsubsection{Problem with the Tully-Fisher Relation}

 Spiral galaxies are observed to show a tight relation 
between luminosity and rotation velocity $V_{\rm obs}$
(which is usually taken to be the maximum rotation velocity).
This Tully-Fisher (TF) relation can be written in the form  
\beq
L=A V_{\rm obs}^{\alpha}\,,
\eeq
where $A$ is the TF amplitude and $\alpha$ the TF slope.
For a given rotation velocity, the observed scatter in $L$
is quite small, typically about 20\% at the bright end.

 Theoretically, the mass of a halo with circular velocity 
$V_{\rm h}$ is given by equation (\ref{MhasVh}).
If we assume the disk mass is $\Md=\md \Mh$, 
we can write
\begin{eqnarray}
\label{TF_Mg}
\Md&=&9.3\times 10^{10} h^{-1}\msun 
\left({\md\over 0.05}\right)\nonumber\\
&&\times
\left({V_{\rm obs}\over 200\kms}\right)^3
\left({\Delta_{\rm h}\Omnow\over 200}\right)^{-1/2}
f_{\rm V}^{-3}\,,
\end{eqnarray}
where $f_{\rm V}=V_{\rm obs}/V_{\rm h}$ is the boost of the observed 
rotation velocity relative to the circular velocity of the halo.
We calculate the boost factor as a function 
of halo concentration $\cs$ and disk mass fraction $\md$,
assuming that the halo response to disk formation 
is through adiabatic contraction  
(see e.g. Mo, Mao \& White 1998 for detailed descriptions).
The disk mass can be converted into a disk luminosity 
by assuming a disk mass-to-light ratio $\Upsilon_d$,
and so equation (\ref{TF_Mg}) can be compared 
with the observed TF relation, once models for 
$\Upsilon_d$ and $f_{\rm V}$ are adopted. Since $f_{\rm V}$ depends 
on $\cs$, the observed TF amplitude is a constraint on $\cs$. 
This is how Mo \& Mao (2000) obtained $\cs\sim 4$.   

 One uncertainty in the predicted TF relation
is the adopted mass-to-light ratio, which is assumed to 
be a constant for a given $V_{\rm obs}$ in the model while in 
reality it may vary from galaxy to galaxy.
Thus, a better way to constrain the theoretical model 
is to use the observed baryon TF relation, which directly
relates the mass of cold gas in a spiral galaxy 
with its rotation velocity $V_{\rm obs}$. Such relation
has been obtained by McGaugh et al. (2000) and 
Bell \& de Jong (2002). By modeling in detail the stellar mass 
and cold gas mass in each galaxy in their sample, 
Bell \& de Jong obtained a baryon mass of  
$\Md\approx 7.0\times 10^{10}\msun$ (assuming $h=0.71$)
for galaxies with $V_{\rm obs}=200\kms$, which is similar to the
value, $\Md\approx 6.4\times 10^{10}\msun$, obtained by   
McGaugh et al. The typical scatter in $\Md$ at the massive 
end is about 20\%. These observed amplitudes can 
be compared directly with the prediction given by
equation (\ref{TF_Mg}), once we know the boost factor 
$f_{\rm V}$ (see e.g. Mo, Mao \& White 1998).
We also assume disks to have exponential mass profiles
and the specific angular momentum is given by a 
spin parameter $\lambda=0.04$. As shown in Mo et al. 
(1998), this is the typical spin parameter required to reproduce 
the observed disk sizes. 

\begin{figure}
\centerline
{\psfig{figure=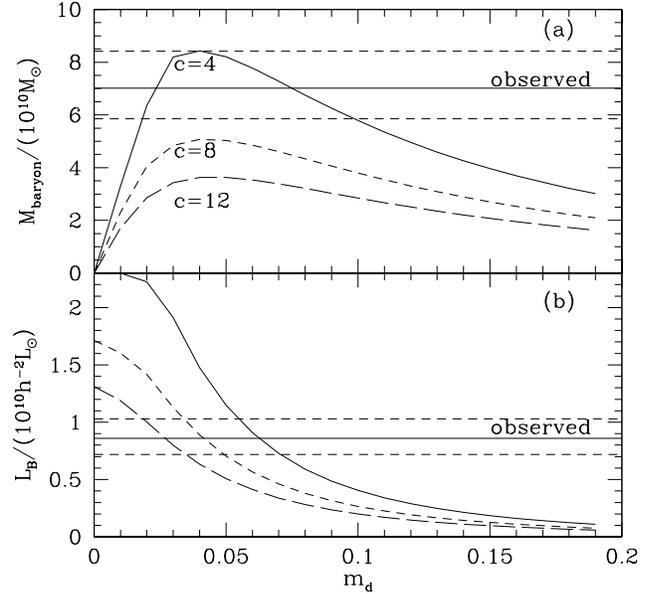,width=0.5\hdsize}}
\caption{(a) The predicted amplitude of the 
baryon TF relation as a function of $\md$ and halo 
concentration $\cs$ (assuming NFW profile) for disk galaxies
with $V_{\rm obs}=200\kms$. The solid horizontal line is the 
mean value observed by Bell \& de Jong (2001), 
while the two dashed horizontal lines represent 20\% 
deviation from the mean.  
(b) The predicted amplitude of the $B$-band
TF relation for $V_{\rm obs}=200\kms$ as required to 
match the observed luminosity function.  
The horizontal solid line is the observed amplitude
based on Tully \& Pearce (2000), while the two dashed horizontal 
lines represent 20\% deviation from the mean.}  
\label{fig_TFandL}
\end{figure}

\begin{figure}
\centerline
{\psfig{figure=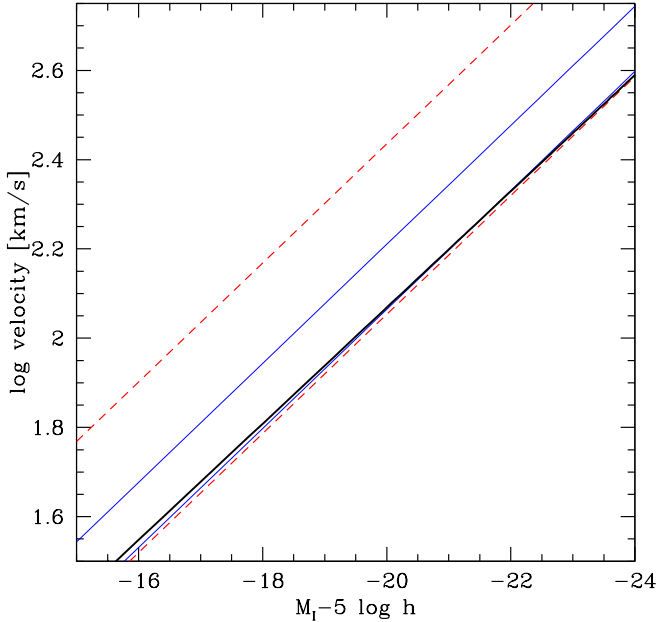,width=0.5\hdsize}}
\caption{
The dashed and thin solid lines are the predicted 
Tully-Fisher relation for $\lambda=0.02$ and 0.04, respectively. 
For each spin parameter $\lambda$, the upper curve shows the 
Tully-Fisher relation where the velocity corresponds to the 
maximum circular velocity of the disk,
while in the lower curve the circular velocity is measured
at $20(V_{\rm h}/220\,{\rm km\,s^{-1}}) h^{-1} \,{\rm kpc}$ (as in
Eke et al. 2001). The thick solid line shows the observed 
Tully-Fisher relation by Giovanelli et al. (1997) in the $I$-band.
}  
\label{fig_tf2}
\end{figure}

Fig.\,\ref{fig_TFandL} shows the predicted TF amplitude at 
$V_{\rm obs}=200\kms$ as a function of $\md$ for several values 
of $c$. A comparison with the observational results 
clearly shows that a low concentration, $\cs\sim 5$, is preferred.
For $\cs=12$, the median value for CDM halos in the 
standard $\Lambda$CDM model, the predicted amplitude 
is lower by a factor of about two.  

 Note that even if a low value of $\cs$ is adopted, 
consistency with observation requires 
$\md$ to be substantially smaller than the universal value $0.16$. 
Thus, only a relatively small fraction of baryons in a 
protogalaxy can manage to settle into the final disk. 
We will come back to the implication of this result in 
subsection \ref{ssec_bf}.

Recently Eke, Navarro \& Steimetz (2001) revisited the zero-point of
the TF relation using high-resolution numerical simulations.
They found that their zero-point is fainter
than the observed value by about 0.5 magnitude for a 
fixed circular speed. They further argued that this
discrepancy is not significant as their simulated galaxies have colours
that are slightly too red compared with the observe TF galaxies.
Their result is therefore inconsistent with what we find here.
In reaching their conclusions, however, they used circular 
velocities measured at a radius given by 
$20(V_{\rm h}/220\,{\rm km\,s^{-1}}) h^{-1} \,{\rm kpc}$, instead of the
maximum circular velocity usually adopted in observations. To show
the effect of their particular choice of the circular velocity, 
we show in Fig.\,\ref{fig_tf2} the TF relation for
$m_{\rm d}=0.16$ and $c=12$; these values are similar to those 
in Eke et al. (2001). We also adopt an $I$-band mass-to-light ratio 
of $1.7h$. Results are shown for two $\lambda$ values (0.02 and 0.04). 
Their simulations likely correspond to the low $\lambda$ value as their
galaxies have quite compact sizes.
With the circular velocity definition adopted by Eke et al. (2001), 
the predicted TF relation matches well the observed TF relation.
This confirms their result. However, if we use
the maximum circular velocity, as in observations, the predicted
TF amplitude is fainter than the observed one by
3.0 and 1.0 mag, respectively. This exercise again highlights that
with a large $m_{\rm d}$ and a large concentration parameter, 
the TF zero-point is difficult to match.

\subsubsection{Problem with the Galaxy Luminosity Function}

 Another constraint on halo concentration 
can be obtained from the observed luminosity function 
of galaxies. Detailed modeling of the luminosity function 
of galaxies in the standard $\Lambda$CDM by Yang et al. 
(2003; see also van den Bosch et al. 2003b)
shows that the halo mass-to-light ratio 
is about $100h$ (in the $B$-band) for halos with masses $\Mh \sim 
10^{12}h^{-1}\msun$ and is higher for both larger 
and smaller masses. This constraint comes from the 
fact that the number density of galaxy-sized halos 
is fixed in the standard $\Lambda$CDM cosmology,
and so the observed luminosity density in the universe
is directly related to the halo mass-to-light ratio. 
Note that this result of the mass-to-light ratio for 
galaxy halos is also consistent with the result obtained
with gravitational lensing (e.g. Hoekstra, Yee \& Gladders 2003).
For a given halo mass-to-light ratio, we can use 
equation (\ref{MhasVh}) to write the 
the luminosity of the galaxy hosted by the halo as 
\begin{eqnarray}
L_B&=&1.86\times 10^{10} h^{-2} \Lsun \left({\Mh/L_B\over 100 h}\right)^{-1}
f_{\rm V}^{-3}\nonumber\\ 
&&\times
\left({\Delta_{\rm h}\Omnow\over 200}\right)^{-1/2}
\left({V_{\rm obs}\over 200\kms}\right)^3\,.
\end{eqnarray}
This has a form similar to the TF relation, with 
amplitude depending 
on the boost factor $f_{\rm V}$. The observed $B$-band TF 
amplitude at $V_{\rm obs}=200\kms$ is $8.6\times 10^{9} 
h^{-2}\Lsun$. This value is quoted without including dust 
extinction correction, because such correction is not made
in the mass-to-light ratio quoted above (see Yang et al. 2003
for detailed discussion). The constraint given by 
this relation is shown in the lower panel of Fig. 1.
To predict a high enough luminosity with $\cs=11$, 
the $\md$ is required to be extremely low, but such 
a low $\md$ is not allowed by the baryon TF relation.
This inconsistency between the Tully-Fisher zero-point and 
the luminosity function in the standard $\Lambda$CDM 
model has also been found in almost all semi-analytical
models of galaxy formation (e.g. Sommerville \& Primack 
1999; Kauffmann et al. 1999; Benson et al. 2002; Mathis et al. 2002; 
van den Bosch, Mo \& Yang 2003a).
As one can see from the figure, consistent results can be 
obtained when $\cs$ is lower than the typical CDM value. 

\subsection {Baryon fraction in spiral galaxies}
\label{ssec_bf}

 Figure \ref{fig_TFandL}
shows that consistency with observation requires 
$\md$ to be much smaller than the universal value $0.16$,
even if a low value of $\cs$ is adopted. Thus, only a small fraction 
of baryons in a protogalaxy can manage to settle into the final disk
(see also the discussion in Salucci \& Persic 1997). 
A similar conclusion can be reached from detailed 
modeling of the mass components in the Milky Way.
Based on the kinematics of Galactic satellites and halo objects,
the halo mass of the Milky Way is estimated to be about
$2\times 10^{12}\msun$ within a radius of about 
$200\kpc$ (Wilkinson \& Evans 1999; Sakamoto et al. 2002).
Within this radius, the mass is still dominated by the Milky 
Way halo, because it is much smaller than the distance
to the other bright galaxy in the Local Group, M31. 
The K-band absolute magnitude of the Milky Way
obtained by Drimmel \& Spergel (2001) is about $-24.02$, corresponding
to $8.3\times 10^{10}{\rm L}_\odot$. Assuming
$(M/L)_K \approx 1$ (e.g., Binney \& Merrifield 1998), we find
a total stellar mass of about $8.3\times 10^{10}\msun$,
which gives a baryon/total mass ratio of $\sim 4.2\%$.
A similar number can be obtained from the local baryonic 
surface density, which is approximately
$50\msun/{\rm pc}^2$ (Kuijken \& Gilmore 1991).
Assuming the Galactic Center is 
8\,kpc away, and taking a disk scale-length of 2.24 kpc (Drimmel \&
Spergel 2001), one obtains a mass
$5.6\times 10^{10}\msun$ for the disk. The bulge/disk
mass ratio is about $1/5$ (Kent et al. 1991), and so the
total mass in the bulge and disk is about 
$6.7\times 10^{10}\msun$, which gives a baryon/total 
mass ratio of $\sim 3.4\%$. Since the mass of baryons in other 
components, such as cold and hot gas, is much smaller, 
the total baryon fraction in the Milky Way is therefore 
much smaller than the universal value $16\%$.

\section{Dynamical effects of the baryon component}

  The problems presented above are not necessarily
the failure of the standard $\Lambda$CDM model because, 
as mentioned earlier, galaxy formation involves many complex
physical processes. One simple assumption
made in the model predictions considered above is  
that the growth of a galaxy disk in the halo centre
is a gentle process, so that the gravitational effect of 
the baryon component can be modeled by adiabatic contraction. 
This may be a good assumption, at least for halos 
with cuspy inner profiles, as shown by numerical simulations 
(e.g. Jesseit, Naab, Burkert 2002).
In addition, the model also makes an implicit 
assumption that the density profiles of dark matter halos,  
within which disks grow,  have the same properties as the halos
in cosmological $N$-body simulations. This may not be a good 
assumption if the formation of a galactic halo contains a 
fast-collapse phase during which the interaction between 
the baryon component and dark matter can significantly 
affect the structure of the halo. In this case, the formation 
of a galaxy disk is in a pre-processed halo that may have 
properties different from a CDM halo. 

Based on this consideration, we propose  
a scenario of galaxy formation that consists of the 
following aspects:
\begin{enumerate}
\item 
    The formation of the core of potential
    through a fast accretion phase, during which 
    gas cools and collapses to form dense clouds that 
    interact dynamically with dark matter particles;
\item
    The loss of orbital energy of dense clouds  
    due to dynamical friction against dark matter particles,
    which increases the kinetic energy of the dark matter
    in the inner region; 
\item
    A phase of rapid star formation in the 
    fast-accretion phase, which causes the loss of a large amount of 
    gas from the halo centre, and causes the inner part of the 
    halo to expand; 
\item 
    The heating of the gas in the protogalaxy region by 
    the starburst in the fast accretion phase;
\item
    Subsequent formation of a galaxy disk in the slow 
    accretion phase by cooling of the preheated gas in the 
    pre-processed halo.
\end{enumerate}

  In this section we focus on the effect of the baryonic 
component on the halo. 

\subsection{Collapse of the baryon component 
and its interaction with dark matter \label{sec_df}}

\begin{figure}
\centerline
{\psfig{figure=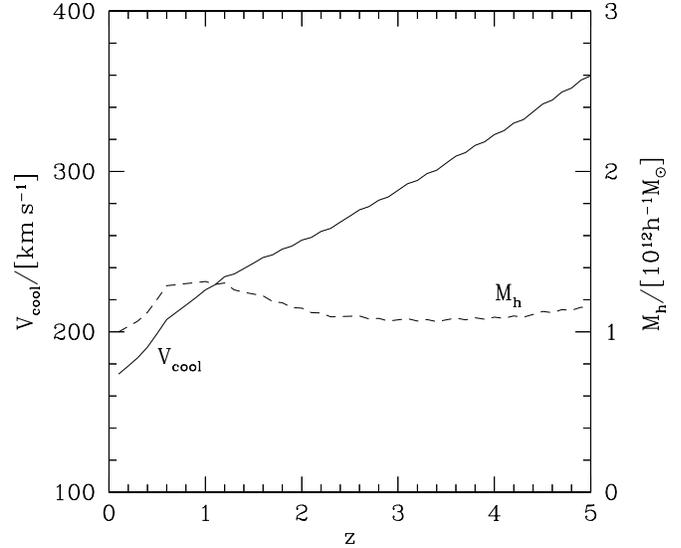,width=0.5\hdsize}}
\caption{The solid curve shows the circular velocity as a function 
of redshift for halos where cooling time is equal to the collapse
time, while the dashed curve shows the masses of such halos.
The cooling time uses the cooling function given by 
Sutherland \& Dopita (1993) for an ionized gas with 
metallicity equal to 0.1 times the solar value.}    
\label{fig_vcool}
\end{figure}

 In the fast collapse phase, i.e. during the formation 
of the core of potential, the interaction between 
the gas component and dark matter is a complicated 
process. In order to gain some insight into this process,
let us first examine some relevant time scales.

 According to Zhao et al. (2003a), the mass of a halo 
in the fast collapse phase increases roughly 
as $M_{\rm h}(z)\propto (1+z)^{-5}$ in the standard 
$\Lambda$CDM model. For galaxy-sized halos
where the fast collapse phases ended at $z\ga 2$,
this relation defines an accretion time scale, 
\beq\label{eq_taccretion}
t_{\rm ac}\equiv M/{\dot M} \sim 1/[5H(z)]\,,
\eeq
where we have used the fact that the scale factor 
increases with time as $a(t)\propto t^{2/3}$
at $z\ga 2$. This time scale is slightly shorter 
than the collapse time scale of the halo, 
\beq
t_{\rm coll}\approx {\pi \rh\over V_{\rm h}}
\sim 1/[3H(z)]\,. 
\eeq
Note that if $V_{\rm h}$ is fixed, then 
$t_{\rm ac}\sim 2/[3H(z)]$. The fact that the 
accretion time scale in the fast collapse regime is 
much shorter than this is because the potential well associated 
with the halo deepens rapidly in this regime, with 
$V_{\rm h}\propto (1+z)^{-5/3}$.

If radiative cooling is effective, then gas
will cool to form clouds before it collapses in the halo 
centre to form stars. For a completely ionized gas,
the cooling time scale can be estimated as
\beq
t_{\rm cool}={3 nKT \over 2 n^2 \Lambda (T)}\,,
\eeq
where $T$ is the virial temperature of the halo, 
$n^2\Lambda (T)$ is the cooling rate per unit 
volume, and $n$ is the mean particle density. We use
the cooling function given by 
Sutherland \& Dopita (1993) for an ionized gas with 
metallicity equal to 0.1 times the solar value.    
This should be compared to the collapse time 
$t_{\rm coll}$.
For halos with $t_{\rm cool}\ll t_{\rm coll}$, gas cannot be 
heated up by the gravitational collapse, while 
for halos with $t_{\rm cool}\gg t_{\rm coll}$, gas may be 
heated to the virial temperature and form a hot halo
which is approximately in hydrostatic equilibrium in the 
dark halo potential well. Fig.~\ref{fig_vcool} shows the 
critical halo circular velocity and mass as a function
of redshift, calculated by equating $t_{\rm cool}$ and 
$t_{\rm coll}$. This mass is quite independent of 
redshift and has a value about $10^{12}h^{-1}\msun$.
Thus, in the absence of heating, radiative cooling is 
effective for halos with masses below that of the 
Milky Way halo at all redshifts.  

  Because of the existence of perturbations on small 
scales, the collapse is expected to be very clumpy in the fast 
collapse phase of the halo.  This, together with
the rapid gas cooling discussed above, implies that the gas 
associated with the collapse of galactic halos
must be in the form of massive cold clouds.
Because of effective cooling, the gas clouds can acquire a high 
binding energy, so as to survive the tidal disruption by the 
dark halo. Once such gas clouds become 
self-gravitating, star formation may occur in situ
before the collpase of the final halo. However, 
based on the results of Kennicutt (1998), only 
about $1.7\%$ of the gas in a galaxy (be it a normal 
spiral or a starburst) is converted into stars during a typical 
rotation time of the galaxy. Thus, the fraction of the gas that 
can be converted into stars during a collapse time of the 
halo is about $0.017 (\pi r_{\rm h})/(2\pi r_{\rm g})$,
where $r_{\rm g}$ is the radius of the galaxy.
Since the gas typically contracts by a factor of 20--30 
before it is supported by angular momentum, only a small 
fraction of the gas can be converted into stars during the 
fast collapse phase. Thus, much of the cold gas can collapse 
towards the halo centre. This is consistent with the fact that 
a lot of cold gas seems to be able to sink to the centre of  
a starburst galaxy to feed the intensive star formation 
there. 

  As the cold clouds sink towards the centre of a dark 
matter halo, they interact with the dark matter through 
gravitation. We can obtain some ideas about the gravitational 
interaction between the cold clouds and dark matter
particles from numerical simulations of major mergers of galaxies 
(e.g. Hernquist 1993). In such simulations, each of the merger 
progenitors is assumed to be a galaxy system consisting
of a disk (sometimes a bulge is also included) embedded 
in an extended dark matter halo. Numerical simulations
show that, when two such systems merge, their halos merge 
first to form a common halo, while the two galaxies 
lose their orbital energy due to dynamical friction,
and sink towards the halo centre to merge. 
Here the dynamical friction plays a key role in the merger of the
galaxies. The situation in our problem is more complicated, 
because the baryon component is still in gaseous form, and 
so its energy can be dissipated through cloud-cloud 
collisions. However, simulations of major mergers
containing gaseous disks (e.g. Barnes 1992)
suggest that such dissipation is important only when 
the gas clouds have already sunk to the centre of the 
halo, while most of the orbital energy of the cold gas 
may be lost through dynamical friction.  In what follows, 
we use simple arguments to gain a qualitative 
understanding of the problem.

According to Chandrasekhar's dynamical friction formula,
the dynamical fiction force on a gas cloud with mass $M$ is
\beq
{\bf F}_M
=-{4\pi G^2 (\ln\Lambda)\rho_{\rm eff}  M^2\over V_M^3}
{\bf V}_M\,,
\eeq
where ${\bf V}_M$ is the velocity of the gas cloud
relative to the dark matter, $\ln\Lambda\sim 10$ is the 
Coulomb logarithm, and $\rho_{\rm eff}$ is the effective
mass density of halo particles producing the dynamical friction.
The rate at which the gas cloud loses its orbital energy 
due to dynamical friction is
\beq
{\dot E}={\bf V}_M\cdot {\bf F}_M\,.
\eeq
For a gas cloud moving in a halo with density profile 
$\rho (r)$, the energy loss rate at radius $r$ 
is approximately
\beq\label{dotE}
{\dot E}\propto M^2\rho (r)/V_M(r)\,.
\eeq
For circular orbits, the decay of the orbital 
radius is given by 
\beq\label{drVM}
{d (rV_M)\over dt}= r F_M/M\propto r\rho M/V_M^2\,.
\eeq
Assuming that the energy loss of the gas cloud
at radius $r$ is deposited in a spherical shell 
with radius $r$
\footnote{This assumption holds 
only as an average for clouds moving on different orbits.}, 
the change in the specific energy of 
dark matter particles at radius $r$ is then
\beq\label{eq_calE}
{\cal E}
\propto {{\dot E} \over 4\pi r^2 \rho(r)}
{dt\over dr}
\propto {M\over r} {\cal G}(r)\,,
\eeq
where
$$
{\cal G}(r)\equiv
{{\overline \rho}(<r)\over\rho(r)}
\left[1+{d\ln V_{\rm h}\over d\ln r}\right]\,,
$$
${\overline \rho}(<r)$ is the mean density 
of dark matter within radius $r$, 
$V_{\rm h}(r)=[G \Mh(<r)/r]^{1/2}$, and the last relation 
uses eqs. (\ref{dotE}) and (\ref{drVM}).

The total energy loss of a cloud is 
\beq  
\Eorb=
\int_0^{\rh} 4\pi \rho r^2{\cal E}(r) dr\,.
\eeq
For an NFW profile, most of the energy loss is in the 
inner region, close to $r=\rs$, where 
$\rho\propto r^{-2}$. 
The energy gain per unit mass by the dark matter 
at radius $r$ is therefore  
\beq
{\cal E}
={\cal E}_{\rm orb} 
{M\over M_0} {\rs\over r}\,,
\eeq
where ${\cal E}_{\rm orb}=\Eorb/M$, and 
\beq
M_0 = {\cal G}^{-1}(r) \rs  \int_0^{\rh} 4\pi r \rho(r) {\cal G}(r) dr\,.
\eeq
Over a large range of $r$, $M_0$ is a constant, and 
$M_0\sim \Mh$ for an NFW profile with $c=4$.

\begin{figure}
\centerline
{\psfig{figure=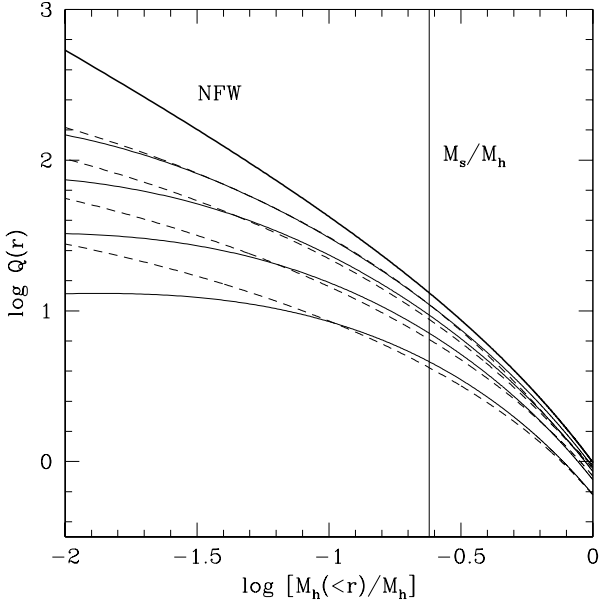,width=0.5\hdsize}}
\caption{The phase-space density profile of dark matter particles.
The thick solid curve show that for the NFW profile, while 
the other solid curves show the profiles taking into 
account the effect of dynamical friction.
Curves from bottom up are results where the 
gas mass is $0.16$, $0.08$, $0.04$ and $0.02$ times
the halo mass. The dashed curves show the results
for the density profile in eq. (\ref{newprofile}) assuming 
$\cc\equiv \rh/\rc=2.2$, $3.5$, $5.5$, $8.0$ (from bottom up). 
The values of $Q$ at the outer radius of the halo
are chosen to match those given by the solid curves.
The vertical line marks the value of $\Ms$ (cf. eq \ref{eq:Ms}).}
\label{fig_Q}
\end{figure}

As one can see, the effect of dynamical friction depends 
on the mass of gas cloud. If we approximate dark halo 
by a singular isothermal sphere, the dynamical friction time 
scale is approximately
$t_{\rm DF} \sim 0.3 f(\epsilon)
t_{\rm coll}\eta/\ln(\eta)$, 
where $\eta\equiv M_{\rm h}/M$ is the ratio between 
the halo mass and the cloud mass, $f(\epsilon)
\sim \epsilon^{0.78}$ describes the dependence on
orbit eccentricity (Lacey \& Cole 1993), and the typical 
value for $f(\epsilon)$ is about $0.5$ according to Tormen 
(1997). Thus, clouds with masses $M\ga M_{\rm h}/20$ can 
lose most of their orbital energy through dynamical 
friction. During the fast collapse, the merger 
progenitors all have comparable mass (major mergers), 
and so dynamical friction is expected to be important 
for most cold clouds. 

 Assuming isotropic velocity dispersion in a halo, 
the phase space density profile can be written as 
\beq
Q(r)={\rho(r)\over [\sigma_i (r)]^3}\,,
\eeq
where $\sigma_i (r)$ is the original velocity dispersion at 
radius $r$. As shown by Taylor \& Navarro (2001),
this phase space density profile can be well approximated by a 
power law $Q(r)\propto r^{-1.825}$ for a NFW density profile.
If we assume the energy transfer from gas clouds at 
a radius $r$ is to be thermalized completely around this 
radius, the effect on dark matter particles is to 
increase their velocity dispersion and reduce the  
phase space density: 
\beq\label{eq:Q_modified}
Q(r)={\rho(r)\over [\sigma_i^2 (r)+\Delta\sigma^2 (r)]^{3/2}}\,,
\eeq
where $\Delta \sigma (r)$ is the increase in the velocity 
dispersion due to dynamical friction.
Using the expression of ${\cal E}$ to obtain 
$\Delta\sigma^2$, we can obtain the phase-space
density profile. Figure \ref{fig_Q}
show the results for different 
values of $f_{\rm gas}\equiv M_{\rm gas}/\Mh$, assuming $\cs=4$. 
As one can see, the phase-space density in the 
inner part of the halo can be significantly reduced,
and the new phase-space density is roughly a constant 
in the central region. The effect is larger for a larger 
value of $f_{\rm gas}$. If $f_{\rm gas}$ is comparable to the 
universal value, $0.16$, more than $20\%$ of the dark 
matter particles will have a roughly constant $Q$.    

 There are a number of uncertainties in the results 
shown above. First of all, although Chandrasekhar's formula 
was shown to describe reasonably well the decay of 
a satellite's orbit in a halo of dark matter particles  
(e.g. Bontekoe \& van Albada 1987; Zaritsky \& White 1988;
Weinberg 1986), the assumption of a local response of the 
halo may not be valid (e.g. Weinberg 1989; Weinberg \& Katz 2002).
Because of this, our assumption that the decay of orbital 
energy of the gas clouds is deposited locally is clearly invalid 
in detail. Unfortunately, as discussed in 
Weinberg \& Katz (2002), a reliable result of the halo response to 
sinking satellites is still beyond the capacity of current 
$N$-body simulations, and so the situation remains unclear. 
We argue, however, that our results are qualitatively 
correct, because the dark matter density is the highest in the inner 
region and so a large fraction of the orbital energy must be 
deposited there. Furthermore, if the mass fraction in cold clouds
is as high as $\sim 10\%$ of the total halo mass, 
the orbital energy of the clouds becomes 
comparable to the total binding energy of the 
dark matter in the inner region at a radius $r\sim 0.1 \rh$. 
Thus, even based on simple energy consideration,  
the `heating' of dark matter by the cold clouds 
is expected to be important in the inner region.

\subsection{Gas outflow and its effect on the halo profile} 

 As the dark matter particles gain energy from gas clouds, 
the dark halo tends to expand to establish a new equilibrium.
On the other hand, the accumulation of gas clouds in the 
halo centre deepens the gravitational potential, which 
can cause the halo to contract. Which effect dominates 
depends on the details of the system in consideration. 
However, the rapid collapse of cold gas in the halo centre 
is likely to be accompanied by rapid star formation, 
which may drive a large amount of the gas out from the halo 
centre. Observations of nearby starbursts indicate 
that the mass-loss rate in a system is typically 1 to 5 
times the corresponding star formation rate (e.g. Martin 1999).
Such mass loss reduces the depth of the potential well
and causes the halo to re-expand, reducing the net 
contraction of the halo due to gas accumulation. 
In the extreme case where all the cold gas that has sunk into the 
halo centre due to dynamical friction can be expelled from the halo 
centre as outflow, the change of the halo profile due to the 
accumulation of gas should be zero (assuming adiabatic process),
and so the only effect that can change the halo profile
is the energy tranfer from the gas clouds due to dynamical 
friction. Thus, in the scenario we are considering here, the net 
effect is always for the dark halo to expand.

\begin{figure}
\centerline
{\psfig{figure=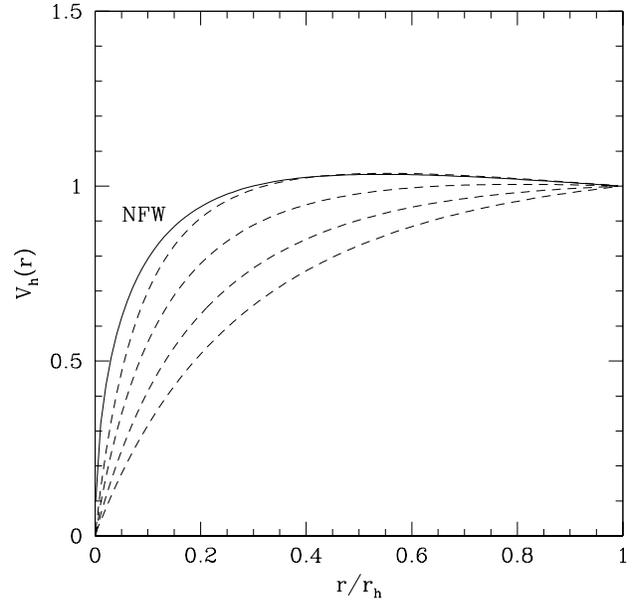,width=0.5\hdsize}}
\caption{The rotation curves of the NFW profile
(solid curve) with $\cs=4$ and of the profile (\ref{newprofile})
with $\cc \equiv \rh/\rc=2.2$, $3.5$, $5.5$, $8.0$ (from bottom up).
The rotation velocity is in units of $V_{\rm h}$ while the radius
is in units of $\rh$.}
\label{fig_Vrot}
\end{figure}

If we assume each mass shell expands adiabatically
after the mass loss, in principle we can solve for the final 
density profile from the hydrostatic equilibrium using phase 
space density profile 
obtained above. Here, we use a simple density model that
approximately reproduces the phase-space density profile.
We assume that the {\it final} density profile can be described by the form: 
\beq\label{newprofile}
\rho(r)={\rho_0 \rc^3\over (\rc+r)^3}\,,
\eeq
where $\rho_0$ is the density at the centre, and $\rc$ 
is a core radius. This profile matches well the NFW profile 
in the outer region. Assuming isotropic velocity dispersion, the  
hydrostatic equilibrium equation is 
\beq
{d(\rho\sigma^2)\over dr}
=-\rho {G [M_{\rm h}(<r) + M_{\rm a}]
\over r^2}\,,
\eeq
where $M_{\rm h}(<r)$ is the halo mass within radius $r$, 
and $M_{\rm a}$ is the added gas mass (i.e. the total gas 
mass collapsed into the halo centre minus 
the gas mass ejected by the outflow).  
Together with the adiabatic equation of state
$\rho(r)/\sigma^3(r)=Q(r_i)$, where $r_i$ is the radius
that encloses the mass $M_{\rm h}(<r)$ in the initial 
density profile, this equation can be integrated to give
\beq
{Q (r_i)\over Q_h}
=\left({\rho\over \rho_h}\right)^{5/2}
\left[1+\int_r^{\rh} {\rho(r)\over\rho_h}
{V_c^2(r)\over\sigma_h^2} {dr\over r}\right]^{-3/2}\,.
\eeq
Here $\rho_h$ and $\sigma_h$ are the density and 
velocity dispersion at the virial radius,   
$Q_h=\rho_h/\sigma_h^3$, and $V_c^2(r)$ is the 
circular velocity of the halo at radius $r$.  
Note that for a given initial density profile, 
$r_i$ is equivalent to $M_{\rm h} (<r)$. With the density profile 
(\ref{newprofile}), and assuming the value of 
$Q_h$ to be the same as that for an NFW profile,
we can obtain the phase-space density profile $Q(r_i)$, or $Q[\Mh(<r)]$. 
The dashed lines in Fig.\,\ref{fig_Q} show the phase-space density 
profiles obtained by assuming different values of
$\cc\equiv \rh/\rc$. For simplicity, we have assumed 
that $M_{\rm a}$ is negligibly small. 
As one can see, over a large range of mass, the phase space 
density profiles obtained 
from \S\ref{sec_df} can be matched by the density
profile (\ref{newprofile}) with appropriate choices
of the value of $\cc\equiv \rh/\rc$. For 
$f_{\rm gas}=0.16$, $0.08$, $0.04$, and $0.02$, we obtain 
$\cc\approx 2.2$, $3.5$, $5.5$ and $8.0$, respectively. 
The density profile (eq. \ref{newprofile}) does not
match the phase-space density in the inner-most part very 
well, which implies the profile corresponding 
to the modified phase-space density distribution 
decreases with radius in the inner-most region. This is 
clearly not physical. In reality, the inner part of the halo 
must relax to an equilibrium configuration that depends on the 
details of the processes involved. However, 
the mismatch involves only a very small fraction of the 
total mass: $\la 2\%$ for an NFW profile with $c=11$
and even smaller for shallower profiles. We therefore 
believe that the simple model adopted here catches the 
essence of the problem.    
 Fig.\,\ref{fig_Vrot} shows the rotation curves 
given by (\ref{newprofile}) with these concentrations,
as compared with the original NFW profile 
with $\cs=4$. Thus, depending on the mass of the gas that
can collapse into the halo centre and ejected by galaxy wind, 
the resulting rotation curve can be flattened substantially
relative to that given by the original NFW profile.  
Within the original scale radius $r_s$, the mass decrease
by dynamical friction is comparable to the total gas mass
in clouds.

\section {Observational Consequences}

\subsection{The Tully-Fisher relation}

 In the slow accretion phase after the formation of the 
core of potential, the virial radius increase roughly 
as $H^{-1}(z)$ without affecting much the inner structure.
The halo concentration will then increase by a factor of 
about $H(\zf)/H_0$ from the formation time to the present time. 
For galaxy halos with typical formation redshift $\sim 2$,
this is a factor of 2 to 3. Thus, for a pre-processed 
halo with concentration $\cc\sim 3.5$ at formation, the 
concentration at the present time will be between 
7 and 10. As shown in Mo \& Mao (2000), disk galaxies formed 
in halos with profile (\ref{newprofile}) and with
such concentrations can match the observed Tully-Fisher relation. 
Thus the flattening of dark matter halos in the fast collapse 
phase by the baryon component can help to reconcile 
the standard $\Lambda$CDM model with the observed 
Tully-Fisher relation. 

\subsection{Dark matter in galaxies}

 The concentration of galaxy halos can be studied by 
looking at the fraction of dark matter mass in galaxies.
The mass in the central parts of galaxies are expected
to be more dominated by dark matter, if dark halos have 
higher concentrations, and vice versa.
Since the total mass can be inferred from observations 
of kinematics, while the contribution from stars 
and gas can be inferred by other means, it is possible
to infer what fraction of the gravitational mass is 
in dark matter. For spiral galaxies, this can be done by
modeling the rotation curves and the stellar mass distribution
in detail.  The results are not yet conclusive 
(see e.g. Bosma 1998 and references therein). 
The rotation curves for some bright spirals are observed to 
decrease just beyond their optical radius, which is best 
explained if the mass in the central region is dominated 
by the stellar component (e.g. Noordermeer et al. 2003). 
On the other hand, there 
are also spiral galaxies, whose rotation curves are quite
flat around the optical radius (e.g. Dutton et al. 2003).  
The rotation curves of many 
bright spiral galaxies can be fitted by the maximal disk model, 
in which the inner rotation curve is assumed to be completely 
dominated by the stellar components.
Although this does not mean that real spirals all have maximal 
disks, it does suggest that the inner rotation curves of some 
bright galaxies may be dominated by the stellar component.
Similar constraints  may be obtained from the fact that the bars 
in some spiral galaxies are fast rotating. This is not expected if 
the central region of a barred galaxy were dominated by 
dark matter, because the bar rotation would be slowed down 
rather rapidly by the dynamical friction of the dark matter
(e.g. Debattista \& Sellwood 2000; 
Weiner, Sellwood \& Williams 2001).
All these observations suggest that some
spirals are dominated by the stellar component in their 
inner regions.   

If the halos of spiral galaxies are
as concentrated as CDM halos given by $N$-body simulations,
most spiral galaxies will be dominated by dark matter, 
unless the disk is unreasonably heavy (e.g. Mo, Mao \& White 
1998). On the other hand, for a pre-processed halo with an 
extended core, the contribution from dark halo 
is significantly reduced, making it more likely to observe 
spirals with inner rotation curves dominated by stellar 
mass.

For elliptical galaxies with moderate luminosities
($L\sim L_*$), recent observations based on the 
kinematics of planetary nebulae show that the observed 
velocity dispersion profiles up to $\sim 5$ effective radii
are consistent with being totally due to stellar mass 
(Romanowsky et al. 2003), indicating the presence of 
little or no dark matter at these radii. As discussed by these authors,
this result is difficult to understand if these galaxies possess 
dark halos with properties similar to those of the CDM halos.
However, if the formation of these galaxies is similar to 
that of the bulges of spiral galaxies, their halos may be 
significantly flattened by the processes envisaged in this paper. 
In this case, the observational result may be easier to understand
with the standard $\Lambda$CDM model.
  
\subsection{Survival of substructures in the halo}

 After the formation of the core of potential, halo
will continue to accrete mass in the form of small halos.
High-resolution $N$-body simulations show that some 
of the accreted halos can survive as subhalos  
(e.g. Klypin et al. 1999; Moore et al. 1999; 
De Lucia et al. 2003). Intriguingly, the predicted
number of sub-haloes in CDM exceeds the observed number of {\it luminous}
satellite galaxies in a Milky-Way type galaxy.
One solution for this crisis may be that some of the
subhalos, especially those of lowest mass, do not form
many stars and hence remain dark. One of the best ways to
detect such substructures is through gravitational lensing. The
``anomalous'' flux ratios in gravitational lenses
are thought to be evidence for substructures (e.g.
Mao \& Schneider 1998; Kochanek \& Dalal 2003 and references therein).

In an NFW halo, most of the subhalos are 
distributed in the outer part, because subhalos can be 
tidally destroyed by tidal force in the inner part.
The situation is different in our scenario, where subhalos may exist   
in the central part of a galaxy halo because of the 
reduced tidal effect in the extended core. 

\begin{figure}
\centerline
{\psfig{figure=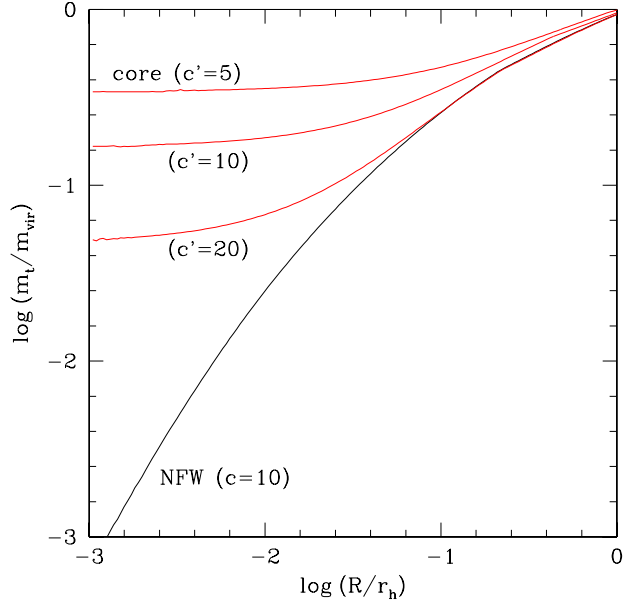,width=0.5\hdsize}}
\caption{The ratio between the mass within the tidal 
radius and the original virial mass for a small halo 
(approximated by an NFW profile with $c=20$) at a given distance from the 
centre of a large halo. The large halo is either 
assumed to have a NFW profile with $c=10$, or has the 
profile given by (\ref{newprofile}), with $\cc \equiv \rh/\rc=5$, 
10, and 20, respectively.}
\label{fig_tidal}
\end{figure}

  Consider the accretion of a minihalo with circular 
velocity $\sim 30 \kms$ by a galaxy halo with a circular 
velocity of $200\kms$. In the presence of a strong UV background, 
minihalos are not expected to trap gas to form stars (e.g. Gnedin 2000),
and so their density profiles are not altered significantly 
by the interaction with the baryonic component. If 
such halos can be approximated as NFW profiles, then according to 
equation (\ref{casM}), they will have a typical concentration $c\sim 20$. We calculate the 
tidal radius of such a minihalo in a galaxy halo 
with or without a core. The tidal radius, $r_t$, of a minihalo 
with mass $m$ moving at a radius $R$ from the centre of a 
large halo with mass $M$ is the minimum of the following 
two radii: (1) a radius at which the gravity of the small halo is equal 
to the tidal force of the big halo, and (2) a radius defined by 
the resonances between the force the minihalo 
exerts on the particle and the tidal force by the 
large halo (e.g. Klypin et al. 1999). The first radius
can be solved from 
\beq
\left({R\over r_t}\right)^3 {m(r_t)\over M(R)}
=2-{R\over M (R)}{\partial M\over\partial R}\,,
\eeq
where $M(R)$ is the mass within radius $R$. The second radius, 
assuming primary resonance, is given by
$\Omega (r)\vert_{\rm small}=\Omega (R)\vert_{\rm large}$, 
where $\Omega$ is the angular speed.
Fig.\,\ref{fig_tidal} shows the mass a minihalo can retain 
within the tidal radius as a function of the distance to 
the centre of the big halo. Results are shown for cases where 
the large halo either has a NFW profile with $c=10$, 
or has the profile given by (\ref{newprofile}) with $\rh/\rc=5$, 
10, and 20, respectively. If the halo has a core 
with $c^\prime\equiv \rh/\rc$ as small as 10, as is required to match the 
Tully-Fisher relation, a minihalo can  
retain more than 20\% of its original mass, 
in strong contrast with the case where the host halo 
has the original NFW profile. This clearly demonstrates 
that the survival of subhaloes in the central part of galaxies 
is likely to be sensitive to the inner mass profiles. 
Note, however, that in Fig.~5, we have neglected the effect 
of baryons which may steepen the central density profiles 
and hence destroy some subhaloes. It is currently unclear
what might be the eventual survival fraction of subhaloes in 
a realistic galaxy. 
  
\section{Preheating and galaxy formation} 

\subsection {Preheating of a protogalaxy region}

 In order for the starburst in the core of potential 
to eject a large amount of gas from the dark halo potential well, 
sufficient number of stars must form so that the kinetic 
energy from supernova explosions from massive stars can heat the gas 
to a temperature comparable to the escape temperature.
For an NFW profile, the escape velocity can be 
calculated from equation (\ref{Phiasr}).
We obtain
\beq
V_{\rm esc}(0)=\sqrt{-2\Phi(0)}=\beta  V_{\rm h}\,,
\eeq
where $\beta \sim 3.1$ and $3.7$ for $c=4$ and 10, respectively.
We write the star formation rate as
\beq
{\dot M}_\star ={M_{\rm gas}\over \tau t_{\rm ff}}\,,
\eeq
where $M_{\rm gas}$ is the total mass of cold gas
that has assembled in the halo centre in 
the fast accretion phase, $t_{\rm ff}$ is the 
free-fall time of the gas, and $\tau$ is a constant.
The total mass of stars that can form can be written as 
\beq
M_\star={\dot M}_\star t_\star =M_{\rm gas} 
\epsilon_\star\,,
\eeq
where $t_\star$ is the star formation time,
and $\epsilon_\star\equiv t_\star/(\tau t_{\rm ff})$
is the star formation efficiency. The total kinetic energy 
from Type II supernova explosions can be written as 
\beq
E_{\rm sn}=M_\star \nu\epsilon_{\rm sn}\,,
\eeq
where $\nu$ is the number of supernovae per unit
mass, and $\epsilon_{\rm sn}$ is the kinetic 
energy per supernova. We adopt 
$\epsilon_{\rm sn}=10^{51}{\rm erg}$
and $\nu=(125\msun)^{-1}$. The value of $\nu$ 
quoted here assumes a Salpeter initial mass function (with
a lower  cutoff of $0.1M_\odot$ and an upper cutoff
of $100M_\odot$).
If a fraction of $\epsilon_0$ of this energy is to heat 
the gas, the specific thermal energy of the gas will be  
\beq
{\cal E}={\epsilon_0 E_{\rm sn}\over M_{\rm gas}}
=\epsilon_0 \nu\epsilon_{\rm sn}\epsilon_\star
\approx (630 \kms)^2 \epsilon_0\epsilon_\star\,.
\eeq
Requiring this specific energy to be equal to the escape 
temperature, we obtain the star formation efficiency,
\beq\label{eq:mustar}
\epsilon_\star = \left({\beta  V_{\rm h}\over 630\kms}\right)^2
\epsilon_0^{-1}\,.
\eeq
Note that this relation is meaningful only
when $\epsilon_\star<1$, and so it only applies for halos
with circular velocities smaller than  
$\epsilon_0^{1/2} \beta ^{-1} 630\kms$.
For larger halos, all cooled gas can form stars 
and so $\epsilon_\star\sim 1$. This scaling relation   
between star formation efficiency and halo circular velocity 
is similar to that obtained by Dekel \& Silk (1986)
based on a calculation of supernova remnants in a uniform 
medium, and has been extensively used in semi-analytic
models of galaxy formation (e.g. White \& Frenk 1991).
The value of $\epsilon_0$ obtained by 
Dekel and Silk is about $0.02$. Since $\epsilon_\star$ must be 
smaller than 1, this low value of $\epsilon_0$
implies that  wind can only escape in small 
halos, with $V_{\rm h}\la 30\kms$ (assuming $\beta =3$).
On the other hand, observations of supernova-driven wind in 
starburst galaxies require much higher efficiency.
The observations of superwind in local starburst galaxies 
suggest that most of the supernova energy is contained 
in the superwind (Heckman et al. 2000). Such high efficiency
may be more relevant to our discussion, as we are mainly 
concerned with starbursts. In this case, wind can escape
from halos with circular velocities as high as 
$200\kms$. In principle, the effective value of 
$\epsilon_0$ can be larger than 1, if active galaxy 
nuclei associated with starbursts also contribute 
to the heating. 

 The gas that has been heated up to the escaping temperature 
will escape the core of potential on a time scale
$t_{\rm esc}\sim r_{\rm cp}/V_{\rm esc}\la 1/[10 \beta  H(\zf)]$, 
where $r_{\rm cp}$ is the radius of the core of potential, 
and $H(\zf)$ is the Hubble constant at the redshift $\zf$
when the core of potential was formed. This time scale 
is shorter than the time scale for mass accretion in the slow 
accretion phase, which is typically the age of the universe.
Thus, the out-going wind will interact with the gas outside 
the core of potential before the gas is accreted by the 
potential well. The average overdensity within a region that 
will eventually form a halo at the present time is 
$\delta (\zf) =\delta_c/D(\zf)$, where $\delta_c\sim 1.68$ is the
critical linear overdensity for spherical collapse, and $D(\zf)$
is the linear grow factor at $\zf$. 
For a flat universe with $\Omnow=0.3$ we have 
$\delta (\zf)\approx 0.43$, $0.53$, $0.71$, $1.03$   
for $\zf=4$, 3, 2 and 1, respectively. 
Thus, for halos with an early formation time, the overdensity 
around the core of potential is moderate at $\zf$.
Based on the extended Press-Schechter formalism
(Bond et al. 1991), one can show that, in the standard 
$\Lambda$CDM model, more than half of the mass 
of a present galaxy-sized halo was in progenitors with circular 
velocities $\la 50\kms$ at redshift $z=2$ -- 3.  
Such progenitors can retain only a small fraction of their gas either 
because their potentials are too shallow to trap gas heated by 
photoionization, or because supernova explosions
in themselves or from nearby galaxies can effectively expel 
the gas (e.g. Scannapieco, Ferrara \& Broadhurst 2000). 
Thus, at the formation redshift of a present-day galaxy 
halo, most of the mass that is not in the core of potential 
is in halos that cannot trap much gas. Thus, the wind 
generated in the core of potential by starburst will be 
propagating in a relatively uniform medium. 

Because the gas density in the post shock medium is 
in general quite low and so radiative cooling of the shocked
gas is negligible, we may use the wind solution of Weaver et al. 
(1977) (which assumes a constant energy injection rate 
and neglects radiative cooling and galaxy potential) 
to get some idea about how much gas can be affected by the wind. 
In this case, the radius of the shock front changes with time as
\beq
R_{\rm s}\sim \left({125\over 154\pi}\right)^{1/5}
\left({E \Delta t ^2 \over\rho}\right)^{1/5}\,,
\eeq
where $E$ is the total supernova energy, $\rho$ is 
the gas density of the medium, and $\Delta t$ is the 
propagation time. Inserting $E=(1/2) M_{\rm wind} \beta ^2 V_{\rm h}^2$ 
into the above expression, and using the relations between
$\Mh$, $\rh$ and $V_{\rm h}$, we obtain 
\beq
R_{\rm s}\sim 10 \rh \left({\Delta t\over t_f}\right)^{2/5}
\left[{f_{\rm wind}\beta^2\over 9}\right]^{1/5}\,,
\eeq
where $t_f \approx 1/H(z)$ is the age of the universe when the starburst
is generated, which is set to equal the formation time 
of the core of potential, and $f_{\rm wind}$ is the 
gas mass fraction in the wind. This radius corresponds to a mass
$M\sim (1000/\Delta_{\rm h})(\Delta t/t_f)^{6/5} \Mh$, where 
$\Mh$ is the mass of the core of potential, i.e. the mass 
of the halo at the formation time. Thus, within a Hubble 
time, the wind can affect a large volume around the halo.    
For a halo that formed at a redshift $\zf$ 
with circular velocity $V_{\rm h}$ and concentration $\cs$,
the total mass at the present time is about 
$H(\zf)/H_0$ times the progenitor mass. Thus, for a 
halo formed at $z\sim 3$, all the gas in the halo proper  
can be affected by the wind generated in the core of 
potential, if the star formation efficiency is as high 
as discussed above. 

  Unfortunately, the details about the 
interaction between the outgoing winds and the intergalactic 
medium are very complicated, and we are not able to make 
a detailed discussion here. If we assume that half of 
the energy of the wind is thermalized in the protogalaxy 
region, then the specific thermal energy gained by the 
gas is
\beq
{\cal E}= [H_0/H(\zf)] (\beta  V_{\rm h})^2/4\,.
\eeq
The corresponding specific entropy of the gas is
\begin{eqnarray}
{\cal S}&\equiv& {T\over n^{2/3}}
={[H_0/H(\zf)] f_{\rm wind} (\beta  V_{\rm h})^2\over 
4[{\overline n}(\zf) (1+\delta_f)]^{2/3}}\nonumber\\
&\sim& 8.5\times 10^2\,{\rm keV\,cm^2}
{[H_0/H(\zf)]\over (1+\zf)^2(1+\delta_f)^{2/3}}\nonumber\\
&&\times
\left({f_{\rm wind}^{1/2}\beta  V_{\rm h}\over 200\kms}\right)^2 
\end{eqnarray}
where ${\overline n} (\zf)$ is the mean gas density 
of the universe at $\zf$, and $\delta_f$ is the 
overdensity of the protogalaxy region. Thus, in 
terms of the value of ${\cal S}$, heating is more 
effective for halos with higher $V_{\rm h}$ and lower $\zf$. 
Taking $\zf=3$, $\delta_f=1$, and $\beta =3$, we have 
${\cal S}\sim 75 (V_{\rm h}/200\kms)^2 {\rm keV\,cm^2}$.
This should be compared with the characteristic specific
entropy of the gas in a virialized halo:
$S_{v}=T_{\rm v}/n_{\rm v}^{2/3}\sim 70 {\rm keV\,cm^2}
(V_{\rm h}/200\kms)^2/(1+z)^2$.  Note that both ${\cal S}$
and ${\cal S}_v$ scale with $V_{\rm h}$ in a similar way, and 
${\cal S}\ga {\cal S}_{\rm v}$ at all $z\ga 0$. 
Thus the heating from the core of potential can have an 
important impact on the subsequent collapse of the gas 
into galaxy halos.   

\subsection {Formation of disk galaxies}

  In the scenario we are considering here, it is quite 
natural to associate the formation of galaxy bulges 
with the fast collapse phase. Depending on whether a 
significant disk can grow in the subsequent slow 
accretion phase, a spiral galaxy with some disk/bulge 
ratio will form. Since the efficiency of gas cooling 
is reduced in a preheated gas, the amount of gas that 
can eventually form a disk can be much lower than that 
implied by the universal baryon fraction (see Mo \& Mao 2002 
for details). Also, the bulge/disk ratio is expected to depend on 
environment, because galaxy halos in high density regions 
may not be able to accrete much gas before they merge into 
large systems where radiative cooling is no 
longer effective. As pointed out in  Mo \& Mao (2002), 
this may be responsible for the observed morphology segregations 
of galaxies. In what follows, we consider several other 
factors that may have played important roles in determining the 
morphological types of galaxies. 

 As shown by equation (\ref{eq:mustar}), 
the star formation efficiency in the fast collapse phase is 
likely to be more efficient in halos with higher
circular velocity. This effect alone would mean that 
bulges are more important in systems with larger $V_{\rm h}$
(or mass). This is consistent with the observational 
trend that later type galaxies generally have lower 
luminosities (e.g. Roberts \& Haynes 1994). 
If the halo is massive enough, this efficient
star formation may heat all the gas around the halo to such 
a high temperature that further cooling of the gas 
is prohibited, making it impossible for the 
formation of a disk component. This might be the reason 
why there appears to be an upper limit on the 
masses of disk galaxies (e.g. Peebles \& Silk 1990).  

In addition to this mass sequence, halo formation history
may also play a role in determining the bulge-to-disk ratio of 
the galaxy that forms in the halo. Although halos of a given
mass have a typical formation time, the dispersion 
among different halos is large for low-mass halos, and so 
there are significant number of present-day galaxy halos
that can have their cores of potential formed only recently
(e.g. Zhao et al., 2003b). Such halos in general have low 
concentration, as discussed above,  but higher specific 
angular momentum (e.g. Maller, Dekel \& Somerville 2002). 
Gas accretion history by such halos is expected to be different 
from that for a halo with early formation. First, because of 
the late formation, the gas associated with the region that 
eventually collapses to form the core of potential may have 
already been heated up by star formation in small progenitors 
before the formation of the core of potential. 
Second, since the characteristic density is lower at lower 
redshift, cooling is less effective at the formation time
for a later formation. Thus, the formation of the core of 
potential of such halos is not expected to be associated with 
rapid formation of cold clouds and with strong starburst.
Such systems are therefore expected to have small bulges
and may be identified as late-type galaxies.  
The halos of such galaxies are therefore expected to 
have lower concentration, and the inner profiles of their
halos are expected to be less affected by the bulge formation.
Because of the late formation, these systems are expected to 
show weaker clustering than normal galaxies 
(e.g. Mo \& White 1996). The latter prediction is consistent 
with the observations that late-type galaxies, such as low 
surface-brightness galaxies, have weaker correlation in space 
than average spiral galaxies (e.g. Mo, McGaugh \& Bothun 1994).

 An interesting question here is whether there are truly  
bulgeless galaxies. The halos of such galaxies 
should not have been pre-processed by the bulge formation
and so should preserve their original profile. 
Any bulgeless galaxies with halos that contain extended 
cores should therefore be considered as serious evidence 
against the CDM cosmogony. As mentioned 
earlier, the rotation curves of many low surface-brightness
galaxies are better fit with halo profiles with cores. 
Whether or not all these galaxies have a significant bulge 
component is still a question of debate. Giant bright low 
surface-brightness (LSB) disk galaxies such as MALIN-1 are known to 
have normal bulges (e.g. Bothun, Impey \& McGaugh 1997)
and so their halos are expected to be significantly 
pre-processed. Early observations in blue bands show that 
many low-luminosity LSBs are bulgeless (see Bothun et al. 1997), 
but recent observations of such galaxies in near infrared bands 
suggest that many of them have detectable bulges in old stars 
and show episodic star formation (e.g. Galaz et al. 2003).
Since these galaxies are associated with shallow potential
wells, the implied amount of early star formation may be 
significant in driving large amount of gas out from halo 
centre, thereby altering the inner profiles of their halos.
Clearly, detailed modeling is required in order to show if
the scenario we are proposing here is consistent with this 
population of galaxies.  

\section{Discussion}

  We have outlined a scenario of galaxy formation
in which the host halos of present-day galaxies may be 
significantly flattened in the inner region by interaction 
with the gas component in an early phase of rapid collapse 
accompanied by starburst. Our scenario is based on recent 
simulation results that such a phase is expected in the current 
$\Lambda$CDM model of structure formation. We have shown that 
this scenario can help to alleviate several vexing problems in 
current theory of galaxy formation. 

 Much further work is required to make quantitative predictions 
with the scenario we have proposed. For example, the treatment 
of dynamical friction is performed in an over-simplified manner. 
In order to model this process quantitatively, high-resolution 
numerical simulations are required. In addition, the
assumption of adiabatic contraction should be more carefully
checked, particularly for cored halo profiles (such as that described by
eq. \ref{newprofile}). Ideally, we should use
cosmological simulations that fully account for
all the relevant processes, such as gas cooling, 
cloud formation, star formation, outflow and dynamical friction.
Unfortunately, such simulations are not yet feasible at the 
present time. A more realistic approach is to tackle the problem 
step by step. For example, the dynamical friction may be 
studied using high-resolution $N$-body simulations where 
cold clouds are represented by simple $N$-body systems;
the interaction between the starburst-driven wind and 
the protogalaxy gas may be studied using controled simulations 
of individual systems. We intend to return to some of these 
issues in future papers.

\label{lastpage}

\end{document}